\documentstyle[12pt]{article}

\def\c{\raisebox{.4ex}{$\chi$}}
\def\d{\delta}
\def\e{\epsilon}
\def\f{\phi}
\def\g{\gamma}

\def\j{\psi}

\def\l{\lambda}
\def\m{\mu}
\def\n{\nu}

\def\ce{{\cal E}}

\def\la{\left}
\def\ra{\right}
\def\pa{\partial}

\def\Hat#1{\rlap{\kern.10em$\widehat{\phantom G}$}#1}
\def\HAt#1{\rlap{\kern.05em$\widehat{\phantom G}$}#1}

\def\cap#1{\rlap{\kern.1em$\widehat{\phantom{G\vrule height.8em}}$}#1{}}
\def\Cap#1{\rlap{\kern.05em$\widehat{\phantom{G\vrule height.8em}}$}#1{}}

\def\leftrightarrowfill{$\mathsurround=0pt \mathord\leftarrow \mkern-6mu
        \cleaders\hbox{$\mkern-2mu \mathord- \mkern-2mu$}\hfill
        \mkern-6mu \mathord\rightarrow$}
\def\overleftrightarrow#1{\vbox{\ialign{##\crcr
        \leftrightarrowfill\crcr\noalign{\kern-1pt\nointerlineskip}
        $\hfil\displaystyle{#1}\hfil$\crcr}}}

\catcode`@=11
\def\underline#1{\relax\ifmmode\@@underline#1\else
        $\@@underline{\hbox{#1}}$\relax\fi}
\catcode`@=12

\def\PRL{Phys. Rev. Lett.\ }

\def\PRD{Phys. Rev. D}

\def\ba{\begin{array}}
\def\ea{\end{array}}
\def\be{\begin{equation}}
\def\ee{\end{equation}}
\def\bdm{\begin{displaymath}}
\def\edm{\end{displaymath}}
\def\bea{\begin{eqnarray}}
\def\eea{\end{eqnarray}}
\def\nl{\nonumber \\}

\def\by{\over}
\def\lbl{\label}
\def\sp{~~~~~}

\def\su#1{\\ \vglue .2in $^{#1}$ {\it Department of Physics\\
          Syracuse University, Syracuse, NY 13244-1130}}

\def\bt#1#2#3#4
        {\begin{titlepage}
         \centerline{\hfill {#1}}
         \centerline{\hfill {#2}}
         \begin{center}\vskip .2in
         {\large\bf {#3}}\\[.3in]
         {\bf {#4}}\\[.3in]
         {\bf ABSTRACT}\\
         \end{center}
         \begin{quotation}}
\def\et
         {\end{quotation}
          \end{titlepage}}

\newcounter{sxn}

\newcounter{axn}

\def\br{\begin{thebibliography}{abc}}
\def\er{\end{thebibliography}}
\def\rf{\bibitem}

\begin{document}
\bt{SU-4240-503}{February 1992}{Non-Abelian Monopoles Coupled to Gravity}{B.
S. Balakrishna and Kameshwar C. Wali \su{}}

A static configuration of point charges held together by the gravitational
attraction is known to be given by the Majumdar-Papapetrou solution in the
Einstein-Maxwell theory. We consider a generalization of this solution to
non-Abelian monopoles of the Yang-Mills Higgs system coupled to gravity. The
solution is governed by an analog of the Bogomol'nyi equations that had played
a central role in the analysis of non-Abelian monopoles.

\et

Under the combined forces of Coulomb repulsion and Newtonian attraction, one
can envisage static configurations of any number of point masses at arbitrary
locations. This happens when the charges, all of the same sign, are equal in
magnitude with the corresponding masses in certain units so that the
Newtonian attraction is balanced by the Coulomb repulsion. It is remarkable
that such a static configuration is allowed in the framework of general
relativity as a solution of the Einstein-Maxwell equations. This was first
shown by Weyl\cite{wxx} assuming axial symmetry and later generalized by
Majumdar and Papapetrou\cite{mpx}. Hartle and Hawking\cite{hhx} have
interpreted this solution as representing an assemblage of extreme
Reissner-Nordstr\"{o}m black holes. Due to the inherent symmetry between the
electric and magnetic fields in the Maxwell theory, one can replace the
electrically charged point masses of the Majumdar-Papapetrou solution by
magnetic monopoles. In some respects, this suggests an analogy with the case of
magnetic monopoles in the framework of spontaneously broken non-Abelian gauge
theories. When the Higgs field responsible for breaking the symmetry belongs to
the adjoint representation of the gauge group, there exists a `no-interaction'
result\cite{opw} for a system of monopoles governed by the Bogomol'nyi
equations\cite{bxx} in the Prasad-Sommerfield limit\cite{psx}. One is led to a
very simple physical interpretation of this result as due to the balance of
Coulomb-type repulsive forces among the magnetic charges and Yukawa-type
attractive forces due to Higgs fields. A natural question that arises is what
happens when such monopoles are coupled to gravity. In this letter, we consider
a type of coupling to gravity that admits static solutions that may be viewed
as a generalization of the Majumdar-Papapetrou solution to a non-Abelian theory
or as an extension of the Bogomol'nyi equations to include gravity. These
solutions represent static configurations of non-Abelian monopoles held in
place by the gravitational forces rather than the Yukawa forces.

The coupling to gravity of the Yang-Mills Higgs system we consider is given by
the following action\cite{con}:
\be
S = \int d^4x\sqrt{-g}\la\{-{1\by 16\pi Gv^2}R\f^2 -{1\by
4}\la(F_{\m\n}^a\ra)^2
+{1\by 2}\la(D_\m\f^a\ra)^2 -{\l\by 4}\la(\f^2-v^2\ra)^2\ra\},     \lbl{act}
\ee
where $R$ is the Ricci scalar, $F^a_{\m\n}$ is the field strength associated
with the gauge field $A^a_\m$ and $\f^a$ is the Higgs field in the adjoint
representation. $D_\m\f^a$ is the gauge covariant derivative of $\f^a$ and
$\f^2$ is a short form for $\f^a\f^a$. When $\f^2=v^2$, the first term gives
the usual Einstein Lagrangian. The rest of the terms represent the standard
Yang-Mills Higgs system. We seek a static solution in this system assuming
$A^a_0 = 0$. We will set our unit of scale to correspond to $4\pi G=1$.

In the absence of gravity, one can obtain a static solution in closed form in
the Prasad-Sommerfield limit ($\l \to 0$) by use of Bogomol'nyi equations. In
this limit, in terms of the `electric' and magnetic fields
\be
E^a_i = D_i\f^a, \sp B^a_i = {1\by 2}\e_{ijk}F^a_{jk},            \lbl{emf}
\ee
one deduces the following inequality for the energy functional:
\bea
\ce &=& {1\by 2}\int d^3x\la[\la(E^a_i\ra)^2 + \la(B^a_i\ra)^2\ra] \nl
&=& {1\by 2}\int d^3x\la(E^a_i\mp B^a_i\ra)\la(E^a_i\mp B^a_i\ra) \pm
\int d^3x E^a_iB^a_i \nl
&\ge& \pm \int d^3x E^a_iB^a_i,                                   \lbl{eeq}
\eea
where the last integral can be written as a surface integral using (\ref{emf})
and the Bianchi identity for $F_{ij}$. Clearly, this becomes an equality only
if the Bogomol'nyi equations $E^a_i = \pm B^a_i$ are satisfied. For the gauge
group SU(2), a spherically symmetric solution to these equations representing
the 't Hooft-Polyakov monopole (or antimonopole if $-$ sign is chosen for
$\pm$) has been found in closed form\cite{bxx,psx}.

We find it remarkable that this system, when coupled to gravity as in
(\ref{act}), still admits monopole solutions with an ansatz that resembles the
Bogomol'nyi equations. This happens when the dimensionless parameter $4\pi
Gv^2$
becomes unity, i.e. in our units $v = 1$. First, motivated by the
Majumdar-Papapetrou solution, we make the following ansatz for the metric:
\be
ds^2 = g_{\m\n}dx^\m dx^\n = V^2dt^2 - {1\by V^2}dx^idx^i,
\ee
where $V$, independent of $t$, is a function of ${x^i}'$s. The components of
the Ricci curvature are then easily computed:
\bea
R_{00} &=& V^3\pa^2V - V^2(\pa V)^2, \sp R_{0i} = 0, \nl
R_{ij} &=& \la[{\pa^2V\by V}-{(\pa V)^2\by V^2}\ra]\d_{ij} - 2{\pa_iV\pa_jV\by
V^2},                                                           \lbl{ric}
\eea
where $\pa$ refers to differentiation along the spatial directions. This
yields for the Ricci scalar the following expression:
\be
R = -2V\pa^2V + 4\la(\pa V\ra)^2 = 2V^3\pa^2\la({1\by V}\ra).   \lbl{ris}
\ee
In the background of the above metric, we wish to minimize the `energy
functional'
\be
\ce = \int d^3x \la({1\by 4V^2}R\f^2 + {1\by 4}V^2F^a_{ij}F^a_{ij} + {1\by
2}D_i\f^aD_i\f^a\ra),
\ee
obtained from the action (\ref{act}) ignoring the $\l-$term for the moment. The
`electric' and magnetic fields are defined to be
\be
E^a_i = {1\by V}D_i\c^a, \sp B^a_i = {V\by 2}\e_{ijk}F^a_{jk},   \lbl{ebs}
\ee
where $\c^a = V\f^a$. It is now straightforward to rewrite the above energy
functional in terms of $E$ and $B$\cite{ext}. The result agrees with the
previous expression in Eq. (\ref{eeq}) and hence obeys the same inequality.
The lower bound is still given by a surface integral since the factors of $V$
coming from the above definitions cancel out in the product $E^a_iB^a_i$.
Again, the equality holds only if the equations $E^a_i = \pm B^a_i$ are
satisfied. In the present case, this implies
\be
F_{ij}^a = \pm{1\by V^2}\e_{ijk}D_k\c^a.                         \lbl{bog}
\ee
This resembles the well studied Bogomol'nyi equations of flat space theories
except for the factor $1/V^2$.

Since the above procedure yields a minimum of the energy functional, the
Euler-Lagrange equations of motion for $A^a_\m$ and $\f^a$ will be
automatically satisfied. It follows from (\ref{act}) that these equations are
\bea
D_i\la(V^2F_{ij}^a\ra) &=& gf^{abc}\f^bD_j\f^c,                \lbl{ef} \\
V^2D_i\la(D_i\f^a\ra) &=& {1\by 2}R\f^a + \l\la(\f^2-v^2\ra)\f^a,
                                                               \lbl{ep}
\eea
where $g$ (not to be confused with the determinant of the metric) is the gauge
coupling constant and ${f^{abc}}'$s are the structure constants of the group.
It is now easily checked, using (\ref{bog}), that Eq. (\ref{ef}) is identically
satisfied. The ansatz (\ref{bog}) requires, as a consequence of the Bianchi
identity for $F_{ij}$, that $\c$ satisfy
\be
D_i\la({1\by V^2}D_i\c^a\ra) = 0.                             \lbl{ec}
\ee
Using this we find that Eq. (\ref{ep}) is also satisfied (again, momentarily
ignoring the $\l-$term) when the Ricci scalar is given by our previous
expression in Eq. (\ref{ris}). Thus the equations of motion for the matter
fields are all automatically satisfied.

The gravitational equations are obtained by varying the action (\ref{act}) with
respect to the metric $g_{\m\n}$. They take the form
\be
R_{\m\n} - {1\by 2}g_{\m\n}R = {2\by \f^2}T_{\m\n},           \lbl{eg}
\ee
where $T_{\m\n}$ is given by
\be
T_{\m\n} = V^2F^a_{\m i}F^a_{\n i} + D_\m\f^aD_\n\f^a + g_{\m\n}\la\{{1\by
4}V^4F^a_{ij}F^a_{ij} + {1\by 2}V^2D_i\f^aD_i\f^a + {\l\by
4}\la(\f^2-v^2\ra)^2\ra\},
\ee
upto terms involving derivatives of $\f^2$. Remarkably, these are all satisfied
when the following holds:
\be
\f^2 = v^2 = 1, \sp {\rm or} \sp \c^2 = V^2.                  \lbl{chs}
\ee
This can be easily verified using Eqs. (\ref{ric}), (\ref{bog}) and
(\ref{ec}). We note that our neglect of the $\l-$term is justified since it
vanishes under this condition. All that is required of the scalar potential is
that it and its first derivative with respect to $\f^2$ vanish at $\f^2=v^2$.

It remains to solve Eq. (\ref{bog}), or (\ref{ec}), to obtain the final
solution. Upto now we have not made any assumption about the gauge group. If we
have U(1), we can drop the indices $a,b,\cdots$. Then $\c^2 = V^2$ implies $\c
= \pm V$ and hence from Eq. (\ref{ec}) we get $\pa^2(1/V) = 0$. The resulting
metric agrees with the well known Majumdar-Papapetrou solution for the
Einstein-Maxwell theory\cite{mpx}. In this case, the scalar field is completely
frozen at its vacuum value. The solution corresponds to a collection of point
monopoles held together by the gravitational attraction.

Also, we have not made any assumption about the number of monopoles or the
symmetry of the solution. There should exist solutions representing static
configurations of non-Abelian monopoles, but unlike those of flat space
theories, such configurations will be held in place by the gravitational forces
rather than the Yukawa forces. This is because the Higgs field is effectively
at its vacuum value as can be seen from (\ref{chs}). Balance of forces requires
that any such solution has, in our units, its total mass equal to its total
monopole charge. Note that the surviving magnetic field at spatial infinity is
\be
B_i = B^a_i\f^a = \pm{1\by V^2}D_i\c^a\c^a = \pm{1\by
2V^2}\pa_i\la(\c^2\ra) = \pm{1\by V}\pa_iV,                    \lbl{smf}
\ee
where we used Eqs. (\ref{ebs}), (\ref{bog}) and (\ref{chs}). Now, at spatial
infinity, $V$ is expected to have the behaviour
\be
V \to 1 - {M\by 4\pi r}, \sp r \to \infty,                     \lbl{V00}
\ee
where $r$ is the radial distance and $M$ is the total mass of the system. Using
this in (\ref{smf}) we obtain the total monopole charge, given by the total
magnetic flux emanating from the system, to be equal to $\pm M$ that agrees in
magnitude with the total mass. One can also get this result from the minimum of
the the energy functional:
\be
\ce = \pm\int d^3x E^a_iB^a_i = \int d^3x E^a_iE^a_i = \int d^3x {1\by
V^2}D_i\c^aD_i\c^a,
\ee
which after partial integration using (\ref{ec}) reduces to a surface integral
to give the total monopole charge.

Next we consider the gauge group SU(2) and specialize to the case of spherical
symmetry. As in the case of the 't Hooft-Polyakov monopole, we start with the
following ansatz for $\c^a$ and $A^a_i$:
\be
\c^a = \mp\hat{r}^aV(r), \sp A^a_i = {1\by gr}\e^{aib}\hat{r}^b\la(1-W(r)\ra),
\ee
where $\hat{r}$ is the unit vector along the radial direction. This gives, for
the electric and magnetic fields defined in (\ref{ebs}),
\bea
E^a_i &=& \mp\d^{ai}{W\by r} \mp \hat{r}^a\hat{r}^i\la({V'\by V}-{W\by
r}\ra), \nl
B^a_i &=& {V\by g}\la\{\d^{ai}{W'\by r} - {\hat{r}^a\hat{r}^i\by
r^2}\la(rW'+1-W^2\ra)\ra\},                              \lbl{flds}
\eea
where prime denotes differentiation with respect to $r$. The Bogomol'nyi
equations $E^a_i = \pm B^a_i$ then lead to the following nonlinear system:
\bea
r^4F'' &=& e^{-2F},                                      \lbl{LEe} \\
{g\by V} &=& {1\by r} + F',                              \lbl{gV}
\eea
where $F$ is defined by $W=e^{-F}/r$. Eq. (\ref{LEe}) is a special case of
the so-called Lane-Emden equations that play a central role in the internal
constitution of stars\cite{cxx}. To our knowledge, it has not been solved
analytically. However, one can show that there exists a class of solutions with
the asymptotic behaviour
\bea
F \to Cr + D&,& \sp r \to \infty,                        \lbl{F1}  \\
F \to -{\rm ln}r&,& \sp r \to 0,                         \lbl{F0}
\eea
where $C>0$ and $D$ are two constants. The expected behaviour of $V$ given by
(\ref{V00}) thus follows from (\ref{gV}) and (\ref{F1}) if $C=g$ and $M
= 4\pi/g$. This is in accordance with our earlier result that the mass and the
monopole charge coincide. However, the behaviour of $V$ as $r\to 0$ is not that
simple. Direct use of the asymptotic behaviour of $F$ gives $V = \infty$. We
need the next correction to this result to come to any conclusion. It is easily
found that the correction to (\ref{F0}) is of the form
\be
A\sqrt{r}~{\rm sin}\la({\sqrt{7}\by 2}{\rm ln}r + \d\ra),
\ee
for some constants $A$ and $\d$. Use of this suggests that $V$ goes to zero as
$\sqrt{r}$, as $r\to 0$. However, it does so in an oscillatory way. The area of
any sphere surrounding the origin, $4\pi r^2/V^2$, vanishes periodically and
tends to zero in this limit. The Ricci scalar given by (\ref{ris}) goes as
$-4V^2/r^2$ and diverges whenever the area vanishes, the geometry being
singular for those values of $r$. What is then the nature of the singularity at
$r=0$ ? In the case of the Majumdar-Papapetrou solution for the metric, the
area tends to a constant and one approaches the event horizon of the black hole
in this limit\cite{hhx}. In our case the origin does have an event horizon,
but with zero surface area. This suggests that the singularity at the origin is
close to being naked. There are no other horizons to dress this singularity
since $1/V$ behaves smoothly, and hence $V$ does not vanish, for finite values
of $r$.

It seems that the above features are specific to $4\pi Gv^2$ being unity and to
get a better understanding, one needs move away from this critical value. We
may test the sensitivity of our conclusions by modifying the equation for $F$
as follows:
\be
r^{4\g}F'' = e^{-2\g F},
\ee
where $\g$ is somehow related to $4\pi Gv^2$ and when unity yields the
critical case discussed earlier. It is now easy to note that the asymptotic
behaviour of $F$ as $r\to \infty$ is unchanged when $\g$ is greater than 1/2,
but for $r\to 0$ one finds
\be
F \to -\la(2-{1\by\g}\ra){\rm ln}r - {1\by 2\g}{\rm ln}\la(2-{1\by\g}\ra), \sp
r \to 0.
\ee
This shows that near origin $V$ approaches zero as $r$,
\be
V  \to {g\g\by 1-\g}r, \sp r \to 0.
\ee
This is what is expected for a black hole with a finite event horizon at the
origin. This sounds attractive, but we have not been able to derive this case
from an action principle.

The constant $D$ of Eq. (\ref{F1}) is not determined by the data at spatial
infinity, leading to a one parameter family of solutions. It appears to be a
measure of the inverse size of the core containing non-Abelian structure. As it
tends to $\infty$ the core collapses to the origin and the solution coincides
with that of Majumdar and Papapetrou for a single monopole. This case, where
$W$ is zero and $1/V$ is $1+1/gr$, can be verified to follow directly from Eqs.
(\ref{flds}). It lies entirely within the U(1) subgroup of SU(2) defined by the
direction of the Higgs field.

To summarize, we have analyzed a type of gravitational coupling to non-Abelian
monopoles and obtained a solution that may be viewed as a generalization of the
Majumdar-Papapetrou solution or as an extension of the Bogomol'nyi equations.
The case of conventional coupling of gravity to non-Abelian monopoles have been
discussed in the literature\cite{nwp} drawing support from numerical
treatments. The advantage of the present approach is that it admits analytical
treatment to a large extent. Perhaps, this is possible only in the critical
case discussed in this letter and moving away from criticality may require
numerical analysis. Secondly, the present approach opens up the possibility of
analyzing multimonopole solutions in the presence of gravity and it will be
interesting if such solutions exist and the known techniques are applicable in
analyzing them. It will also be interesting if there exists a stationary
generalization analogous to the one that exists for the Majumdar-Papapetrou
solution\cite{iwp}.

We wish to acknowledge Professors S. Chandrasekhar, James Hartle and Nina Byers
for useful discussions. This work was supported in part by DOE under contract
number DE-FG02-85ER40231.

\br
\rf{wxx} H. Weyl, Ann. Physik 54, 117 (1917).
\rf{mpx} S. D. Majumdar, Phys. Rev. 72, 390 (1947); A. Papapetrou, Proc. Roy.
Irish Acad., Ser. A 51, 191 (1947).
\rf{hhx} J. B. Hartle and S. W. Hawking, Comm. Math. Phys. 26, 87 (1972).
\rf{opw} L. O'Raifeartaigh, S. Y. Park and K. C. Wali, \PRD 20, 1941 (1979).
\rf{bxx} E. B. Bogomol'nyi, Soviet Journal of Nucl. Phys. 24, 449 (1976).
\rf{psx} M. K. Prasad and C. M. Sommerfield, \PRL 35, 760 (1975).
\rf{con} In our conventions, $g_{\m\n}$ has the signature $(+---)$. Indices
$\m,\n,\cdots$ run from 0 to 3 and $i,j,\cdots$ from 1 to 3. Indices
$a,b,\cdots$ run over the adjoint representation of the gauge group. Repeated
indices are assumed to be summed.
\rf{ext} According to G. W. Gibbons and S. W. Hawking, \PRD 15, 2752 (1977),
one should add to the gravitational action a surface integral of the trace of
the extrinsic curvature. In our case, we need to include this surface term
(times $\f^2$) in the action (\ref{act}) since the energy functional, as we
defined it, will not otherwise yield the usual result for the total energy of
the system. This surface term cancels a similar term that arises when rewriting
the energy functional in terms of $E$ and $B$.
\rf{cxx} Eq. (\ref{LEe}) arises when one considers an isothermal gas sphere in
gravitational equilibrium. For a detailed discussion of this and related
equations see `An Introduction to the Study of Stellar Structure', S.
Chandrasekhar, Dover Publications (1957), chapter IV. The present equation
agrees with Eq. (383) in this chapter after the identifications $x=r$ and
$\j=2F-{\rm ln}2$. We do not, however, agree with the asymptotic behaviour
quoted in Eq. (449) to be compared with Eq. (\ref{F1}) of this letter.
\rf{nwp} P. van Nieuwenhuizen, D. Wilkinson and M. J. Perry, \PRD 13, 778
(1976). For recent work, see K. Lee, V. P. Nair and E. J. Weinberg, Columbia
and Fermilab preprint CU-TP-539/FERMILAB-Pub-91/312-A\&T; P. Breitenlohner, P.
Forg\'{a}cs and D. Maison, Max-Planck Institute preprint MPI-Ph/91-91.
\rf{iwp} Z. Perj\'{e}s, \PRL 27, 1668 (1971); W. Israel and G. A. Wilson, J.
Math. Phys. 13, 865 (1972).
\er

\end{document}